\documentclass{Interspeech2024}
\usepackage{multirow}

\interspeechcameraready 

\title{Speak in the Scene: Diffusion-based Acoustic Scene Transfer\\ toward Immersive Speech Generation} 

\name[affiliation={1,2}]{Miseul}{Kim}
\name[affiliation={1}]{Soo-Whan}{Chung}
\name[affiliation={1}]{Youna}{Ji}
\name[affiliation={2}]{Hong-Goo}{Kang}
\name[affiliation={1}]{Min-Seok}{Choi}

\address{
  $^1$Naver Cloud, South Korea, $^2$Yonsei University, South Korea
\email{ miseul4345@dsp.yonsei.ac.kr, soowhan.chung@navercorp.com}
    }
\keywords{audio/speech generation, acoustic scene transfer, latent diffusion}

\usepackage{overpic}
\usepackage{enumitem} 
\usepackage{overpic} 
\usepackage{color}

\usepackage{pifont}
\usepackage{amsfonts}
\usepackage{booktabs}
\usepackage{tabularx}

\usepackage{graphicx}
\usepackage{subfig}
\usepackage{pgfplots, mathtools}
\usepackage{xcolor}

\definecolor{turquoise}{cmyk}{0.65,0,0.1,0.3}
\definecolor{purple}{rgb}{0.65,0,0.65}
\definecolor{dark_green}{rgb}{0, 0.5, 0}
\definecolor{orange}{rgb}{0.8, 0.6, 0.2}
\definecolor{red}{rgb}{0.8, 0.2, 0.2}
\definecolor{darkred}{rgb}{0.6, 0.1, 0.05}
\definecolor{blueish}{rgb}{0.0, 0.3, .6}
\definecolor{light_gray}{rgb}{0.7, 0.7, .7}
\definecolor{pink}{rgb}{1, 0, 1}
\definecolor{greyblue}{rgb}{0.25, 0.25, 1}

\newcommand{\PreserveBackslash}[1]{\let\temp=\\#1\let\\=\temp}
\newcolumntype{C}[1]{>{\PreserveBackslash\centering}p{#1}}
\newcolumntype{R}[1]{>{\PreserveBackslash\raggedleft}p{#1}}
\newcolumntype{L}[1]{>{\PreserveBackslash\raggedright}p{#1}}

\newcommand{\Figure}[1]{Figure~\ref{fig:#1}}

\newcommand{\Table}[1]{Table~\ref{tab:#1}}

\usepackage{blindtext}

\renewcommand{\paragraph}[1]{\vspace{1pt}\noindent\textbf{#1}\textbf{.}}
\newcommand{\scriptnote}[1]{\footnote{\scriptsize{#1}}}

\newcommand{\fullname}{Acoustic Scene Transfer Latent Diffusion Model}
\newcommand{\shortname}{AST-LDM}

\newcommand{\furl}[1]{\scriptnote{\url{#1}}}

\newcommand{\xubsection}[1]{\vspace{-5pt}\subsection{#1}\vspace{-3pt}}
\newcommand{\xection}[1]{\vspace{-3pt}\section{#1}\vspace{-3pt}}

\newcommand{\xcaption}[1]{\caption{#1}\vspace{-8pt}}

\begin{document}

\maketitle

\begin{abstract}

This paper introduces a novel task in generative speech processing, Acoustic Scene Transfer (AST), which aims to transfer acoustic scenes of speech signals to diverse environments.
AST promises an immersive experience in speech perception by adapting the acoustic scene behind speech signals to desired environments.
We propose AST-LDM for the AST task, which generates speech signals accompanied by the target acoustic scene of the reference prompt.
Specifically, AST-LDM is a latent diffusion model conditioned by CLAP embeddings that describe target acoustic scenes in either audio or text modalities.
The contributions of this paper include introducing the AST task and implementing its baseline model.
For AST-LDM, we emphasize its core framework, which is to preserve the input speech and generate audio consistently with both the given speech and the target acoustic environment.
Experiments, including objective and subjective tests, validate the feasibility and efficacy of our approach.

\end{abstract}

\section{Introduction}

In our daily lives, we encounter a plethora of environmental sounds or acoustic scenes, including background noise, reverberation, and other auditory stimuli such as chirping birds, laughing children, or the murmur of distant conversations.
The acoustic environment plays an essential role in human interactions, infusing conversations with vitality and often serving as a contextual cue for speech.
For example, we often include supplementary sounds when conversing with others or reading a book to improve comprehension of the spoken content.
Therefore, people often edit recorded speech with various sound samples, creating an immersive experience akin to speech recorded in a real environment.

Remarkable advances in the field of generative models enable the generation of high-quality speech and audio.
Early generative methods, such as variational autoencoders (VAE)~\cite{akuzawa2018expressive,kim2021conditional} and generative adversarial networks (GAN)~\cite{yamamoto2020parallel,kong2020hifi}, have successfully produced natural-sounding speech signals, surpassing the capabilities of previous discriminative methods.
However, they often struggle with instability during training, which can lead to unpredictable and unsatisfactory outputs.
Recent diffusion-based models have effectively addressed these challenges, markedly raising generation performance and achieving remarkable success in the extensive domains of audio and speech.
Some studies~\cite{kong2020diffwave,lee2023facetts,shen2024naturalspeech} have underscored its efficacy in modeling the distribution of speech signals, especially in facilitating multi-speaker speech synthesis as well as high-quality synthesis.
In the realm of music generation, diffusion has demonstrated its proficiency in crafting musical notes across various instruments such as piano, guitar, beats, and even vocals~\cite{hawthorne2022multi,chen2024musicldm,novack2024ditto}.
Furthermore, diffusion models can even generate general sounds by learning auditory characteristics from text description prompts and producing sound effects and foley sounds~\cite{kreuk2022audiogen,yi2023latent,luo2024diff,evans2024stableaudio}.
Notably, AudioLDM~\cite{liu2023audioldm,liu2023audioldm2} demonstrates exceptional performance in audio generation, showcasing its ability to generate audio signals based on CLAP model~\cite{elizalde2023clap,wu2023large}, retrieving acoustic characteristics from reference audio or text captions.
VoiceLDM~\cite{lee2023voiceldm}, a text-to-speech model, is built upon the framework of AudioLDM, and it generates speech signals with prompts and transcriptions to control speech, voice and various speaking environments.
Similarly, DiffRENT~\cite{im2024diffrent} aims to transfer acoustic characteristics such as ambient noise and reverberation from reference speech to input speech signals.

\begin{figure*}
    \centering
    \includegraphics[width=0.95\linewidth]{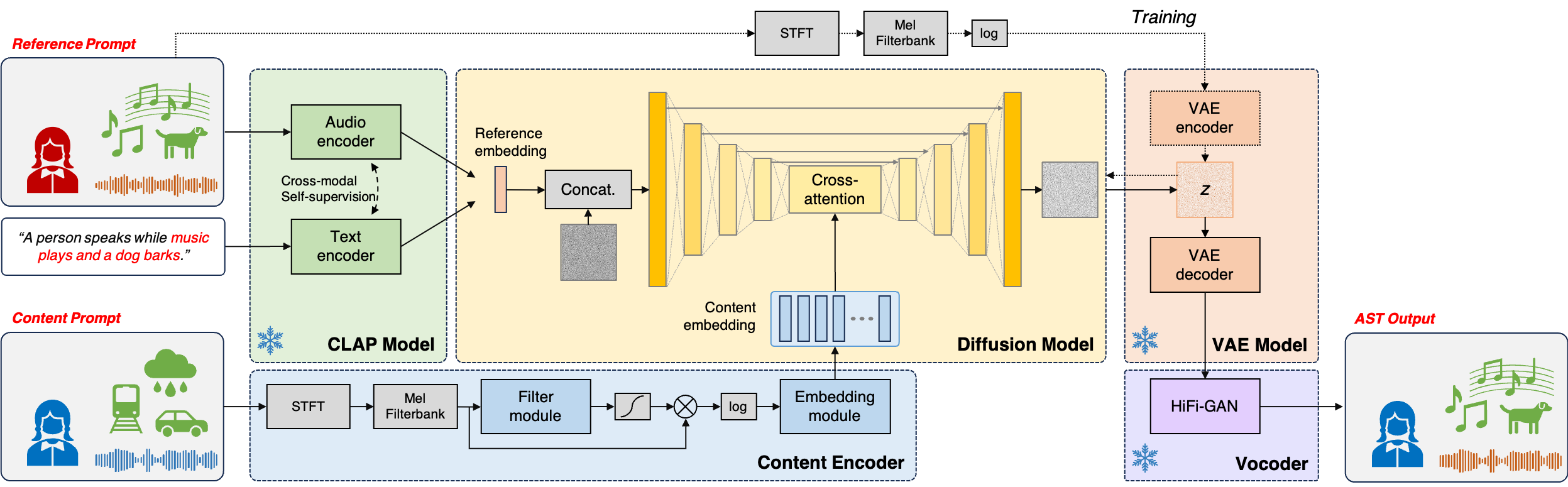}
    \vspace{-5pt}
    \caption{Overview of the proposed model, \shortname. Dotted lines represent the training phase, while solid lines denote the inference process. Blocks marked with \includegraphics[height=0.7em]{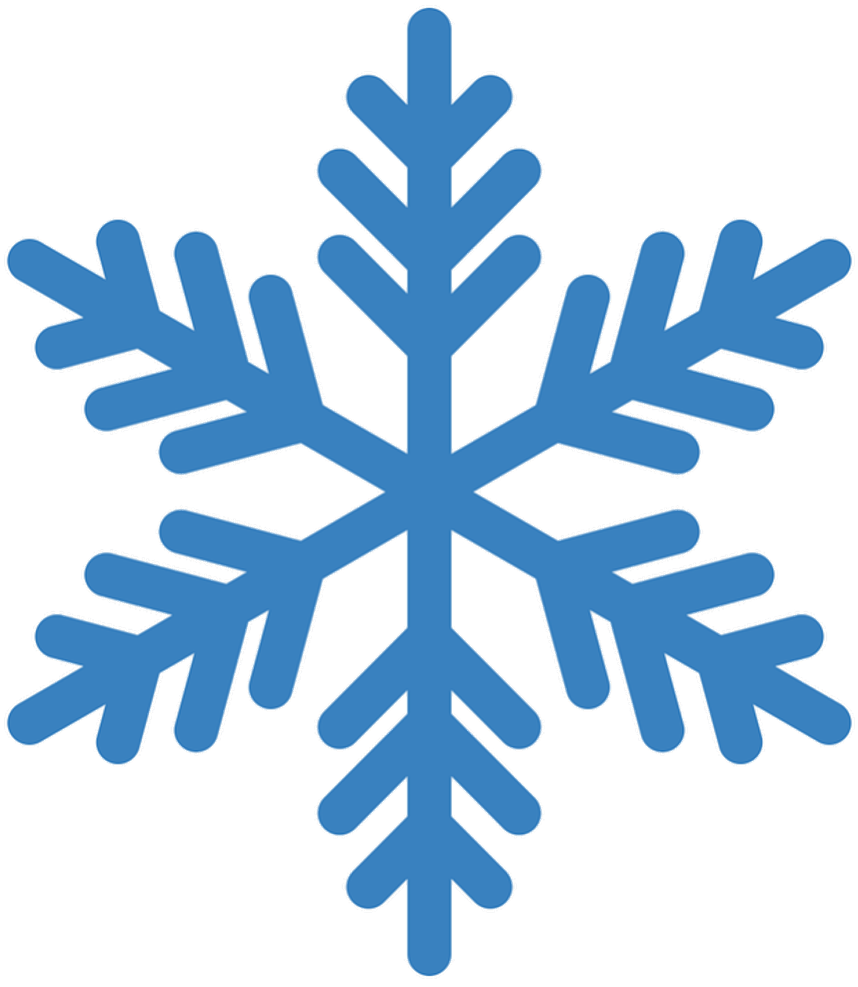} indicate pre-trained components that are not updated during training.}
    \label{fig:astldm}
    \vspace{-7pt}
\end{figure*}

In this paper, motivated by recent developments, we propose a novel speech processing task, \textbf{Acoustic Scene Transfer~(AST)}, aimed at transferring the target acoustic scene onto an input speech signal.
Although the framework of AST is similar to the foundational knowledge presented in DiffRENT, we further generalize the paradigm to encompass general audio rather than focusing solely on ambient noise.
To the best of our knowledge, this is the first time such a task has been proposed for immersive speech generation, and we expect that its potential benefits across various speech processing fields.
While this task may appear similar to adding background sound or room impulse responses as speech augmentation, AST does not simply replicate the acoustic scene of reference signals.
Instead, it re-generates acoustic scenes, allowing for flexibility and diversity on generated scenes and relieving limitations imposed by noise or room impulse response in datasets.
Therefore, AST would encourage more impactful and immersive speech generation beyond conventional natural voice generation.

We implement a latent diffusion model (LDM) for AST, \textbf{\fullname~(\shortname)}, which predicts latent embeddings of a pre-trained variational autoencoder, following AudioLDM and VoiceLDM.
Our proposed model mimics the background sound and acoustic characteristics of the target environment while retaining the spoken term and speaker identity present in the input speech.
\shortname~not only transfers acoustic backgrounds but also facilitates transitions from anechoic to noisy or reverberant environments, and vice versa.
Furthermore, we demonstrate that \shortname~can be conditioned not only by reference audios, as in DiffRENT, but also by text descriptions of acoustic scenes with a help of CLAP embeddings.
By leveraging text prompts, the proposed model gains larger diversity and enriched representations of acoustic scenes, enabling it to generate unseen acoustic situations.
\xection{Diffusion-based Audio Generation}
\vspace{3pt}

Diffusion~\cite{ho2020denoising,song2020score} leads the field of generative learning by predicting the data distribution from Gaussian noise, and it allows the model to generate outputs with a large variability.
The advent of latent diffusion~\cite{rombach2022high}, which operates in a latent space, has reduced excessive computational costs for processing raw data and increased the applicability of diffusion framework on various domains.
In audio generation, diffusion models have achieved significant breakthroughs, as demonstrated in their capability in vision modality.
Its application in speech synthesis led to the high-quality generation of speech waveforms conditioned by mel-spectrograms~\cite{chen2020wavegrad}.
Beyond speech synthesis, diffusion techniques have been applied in speech enhancement~\cite{welker22speech,lu2022conditional} and separation~\cite{scheibler2023diffusion,lutati2023separate}, enabling the generation of clear speech from noisy inputs.
Moreover, diffusion has demonstrated its versatility in creating a broad spectrum of audio content including natural sounds and instrumental music.

The method most closely aligned with ours is AudioLDM, which generates audio sounds by utilizing CLAP model to extract embeddings for the conditional generation.
It adopts a self-supervised training approach; the model is trained to predict pre-trained latent embeddings for a broad spectrum of audio sounds, with the CLAP embedding of the audio itself serving as a reference prompt.
During inference, the model employs the text encoder of CLAP model, using text descriptions of sound to steer the generation process rather than relying on the audio prompt.
It is enabled by the common latent space between the audio and text encoders of CLAP model, which benefits from its training in cross-modal self-supervision.
This approach is particularly effective in generating a varied range of audio outputs, including the ability to generate unseen audio outputs that are not present during the training stage.
Inspired by AudioLDM, VoiceLDM~\cite{lee2023voiceldm} focuses on text-to-speech (TTS) synthesis, incorporating both text transcriptions and audio signals as conditions for training.
It not only focuses on synthesizing clean speech but also simulates speech within the described environment.
Leveraging CLAP model, VoiceLDM adjusts the acoustic scene based on textual scene descriptions, employing a dual classifier-free guidance method~\cite{ho2021classifier} to balance the influence of both conditions.
DiffRENT~\cite{im2024diffrent} presents another innovative method, focusing on the transfer of acoustic scenes from reference to target speech.
Based on a standard diffusion model, it directly estimates the target mel-spectrogram, employing an environmental encoder instead of CLAP model.
The environmental encoder is jointly trained with the diffusion model, sharing the same training criteria.
DiffRENT changes the ambient noise and room acoustics of the target speech to match those of the reference while preserving the spoken content.

In this work, we draw upon the training and conditioning strategies as well as frameworks of these models to generate optimal latent embeddings for AST.
Our work aims to extend the boundaries of audio generation, leveraging the capabilities of diffusion models to create immersive speech experiences.

\xection{\shortname}

In this section, we introduce our proposed model, \shortname, and outline its training strategy for the acoustic scene transfer task. 
The overall structure of \shortname~is given in~\Figure{astldm}.
\shortname~is conditioned on two prompts: the reference prompt, which consists of the target environments, and the content prompt, representing the input speech signal aimed at preserving the speaking contents in the output signal.
Our model is inspired by the structure of AudioLDM which includes the diffusion model, CLAP encoder, VAE, and HiFi-GAN~\cite{kong2020hifi}.
The VAE is charged with extracting the target latent embedding from the mel-spectrogram, while HiFi-GAN serves as the vocoder, converting the mel-spectrogram into an audio waveform.
In the following subsections, we will offer a detailed demonstration of the modules of our proposed model.

\xubsection{Reference Prompt}
In training stage, the reference prompt comprises speech and its speaking environments, including background sounds and reverberation.
The audio encoder of CLAP model processes the reference prompt to obtain the representation of acoustic scenes.
Previous studies on AudioLDM and VoiceLDM have demonstrated the capability of generating audio using CLAP embeddings, suggesting their effectiveness in capturing both the acoustic environments and speaking styles of the audio prompts.
In the inference stage, we utilize text descriptions of the target acoustic scenes in conjunction with the text encoder of CLAP model, in addition to the reference audio prompt used during training.
Since CLAP model is pre-trained using cross-modal self-supervision, we expect that the text description can effectively replace the audio prompt.
Utilizing a text prompt offers the potential for greater diversity and enriched representations of acoustic scenes, enabling the model to generate scenarios that have not been encountered before.
In our experiments, we simulate reference prompts as noisy, reverberant, or both noisy and reverberant scenes, and also clean speech signals to be transferred into a clean environment.

\xubsection{Content Prompt}
The content prompt consists of the target spoken term that is intended to be retained in outputs of \shortname.
Since our goal is to transfer acoustic scenes across diverse environments, the content prompt may contain environmental sounds behind speech, and these acoustic elements should not remain in output signals.
Consequently, it is essential to precisely preserve the speech signal from the content prompt and condition it onto the diffusion model.
To achieve this, we designed a content encoder based on a transformer structure, consisting of two cascaded modules.
The content prompt goes into the filter module, which returns a ratio mask using a sigmoid function to be multiplied by the mel-spectrogram of the content prompt.
The filter module removes undesired acoustic components, ensuring stable conditioning.
Subsequently, the masked mel-spectrogram is passed to the embedding module to encode the content prompts onto the latent space.
Like the reference speech, we simulate the content prompt using noisy, reverberant, both noisy and reverberant, and clean speech signals.

\xubsection{Latent Diffusion}
Given the reference and content embeddings as conditions, the latent diffusion model is trained to predict the target audio embedding in the latent domain.
As mentioned, the latent embedding of the target audio is extracted using the pre-trained VAE model, which compresses mel-spectrograms into stochastic representations.
We utilized the same U-Net structure used in AudioLDM for the diffusion model.
The reference embedding is repeated along the time axis and concatenated with the input of the diffusion model, while the content embedding is conditioned using the cross-attention method on the bottleneck feature of the U-Net.
The target audio is simulated by utilizing the speech signal of the content prompt and the acoustic scenes of the reference prompt.
During the training stage, 
we used the same speech signal in both reference and content prompts but simulated with different acoustic scenes, and the simulated reference prompt also served as the target audio signal.
\vspace{2pt}
\xection{Experimental Settings}
\vspace{3pt}

\paragraph{Data Preparation}
For training, we utilized 960 hours of speech data from LibriSpeech~\cite{panayotov2015librispeech} corpus and curated additional audio samples from AudioCaps~\cite{kim2019audiocaps}, a balanced subset of AudioSet~\cite{gemmeke2017audio}, and VGGSound~\cite{chen2020vggsound} to provide background sounds.
Room acoustics were simulated using room impulse responses (RIRs) from DNS-Challenge dataset~\cite{reddy2021dns}, consisting of both simulated and real-recorded samples.
We divided the RIR dataset into two subsets: two simulated room impulse responses from each room size and RWCP real impulse responses for evaluation, with the remaining data allocated for training.

For evaluation, we employed both audio samples and text descriptions for the reference prompt.
Audio samples were obtained from the \textit{test-clean} subset of the LibriSpeech corpus for speech utterances, the \textit{eval} set of AudioSet, and the RIR evaluation set of DNS-Challenge dataset.
We prepared 1,800 speech samples from LibriSpeech, half for clean content and the rest for clean reference prompts.
Since we used same speech signal for reference and content prompts during training, we aligned the genders of speakers within reference prompts and content prompts.
We overlaid acoustic scenes onto the clean speech samples to create content and reference prompts.
In the case of using text descriptions for reference prompts, we generated 1,200 sentences\scriptnote{Samples available: \url{https://ast-ldm.github.io/demo}} using GPT4~\cite{achiam2023gpt4}, each describing an acoustic scene with the format:
\textit{`A [male/female] speaks in [place] with [background sounds] behind.'}
For noiseless scenes, we used the text prompt: \textit{`A [male/female] speaks in a quiet room'}.

In addition, to compare with AudioLDM and VoiceLDM, we leveraged audio samples of LibriSpeech and AudioCaps.
Since AudioCaps involves text captions of audio samples, we used them as a reference text prompt, as well as a reference audio prompt, and transcriptions of LibriSpeech, were used for the text input of VoiceLDM while we used corresponding speech signals for \shortname.
We also included real-world recordings from VCTK-DDS dataset, featuring clean-noisy speech pairs captured in real environments with ambient noise and reverberation.
For VCTK-DDS dataset~\cite{li2021dds}, we retained audio samples without adding additional background sounds or reverberation, using them as target audio examples.
Following the evaluation data ontology of DiffRENT, we selected 2 speakers (p234, p241) and a recording environment, \textit{livingroom1 in Uber}.
To ensure consistency across datasets, audio samples were resampled to 16kHz and segmented into 10-second chunks, with shorter samples padded to fit the duration.
For the simulation, we mixed speech and environmental sound with a random signal-to-noise ratio (SNR) ranging from 4 to 20 dB.

\paragraph{Model Structure}
The architectural configurations of the diffusion model, CLAP model, VAE model, and HiFi-GAN remain consistent with those described in~\cite{liu2023audioldm,lee2023voiceldm}.
For the content encoder, we designed it with 2 transformer layers for the filter module and 2 convolution layers followed by 4 transformer layers for the embedding module.
Each transformer layer utilizes 256 hidden dimensions with 8 attention heads.
Reference prompts, content prompts, and target speech signals are transformed into 64-dimensional mel-spectrograms on a logarithmic scale, extracted every 10ms with a frame length of 64ms.

\paragraph{Training Strategy}
We adopt a training scheme similar to that of AudioLDM and VoiceLDM and introduce the dual classifier-free guidance method~\cite{ho2021classifier} to train our model conditioned by two different prompts.
During training, we randomly dropped reference and content embeddings with probabilities of 0.1 and 0.1, respectively.
The model is trained using L2 loss to estimate added noise during the diffusion forward process.
While the VAE, CLAP, and HiFi-GAN are pre-trained and left frozen, the diffusion model and content encoder are jointly trained.

\paragraph{Evaluation Protocols} 
During inference, we used a DDIM sampler with 100 sampling steps, setting 1.0 for both dual classifier-free guidance strengths. 
We measured performance in terms of scene similarity and content preservation.
Scene similarity was assessed using CLAP-based Fr\'echet Audio Distance (FAD) and CLAP similarity metrics.
CLAP similarity was calculated using both the audio encoder (CLAP$_\mathcal{A}$) and, if available, the text encoder (CLAP$_\mathcal{T}$).
Content preservation was evaluated using word-error-rate (WER) employing Whisper-small.en~\cite{radford2023robust} model.
Also, speaker similarity (SSM) between content prompts and output signals was measured using the embedding distance of RawNet3~\cite{jung2022pushing} model.
For the subjective evaluation, we conducted listening tests with 17 listeners to assess the overall perceptual quality of the generated audio samples. 
The assessment of audio quality was conducted using a 5-point mean-opinion-score (MOS) scale, where scores ranged from 1 (bad) to 5 (excellent).
Lower scores for FAD and WER indicate better performance, while higher scores for CLAP, SSM and MOS suggest better performance.

\begin{table}[t]
\centering
\footnotesize
\xcaption{Objective evaluation of generated speech with simulated dataset conditioned on audio reference prompts ($\mathcal{R}$: $\mathcal{A}$).}
\begin{tabular}{l|ccccc}
\toprule
\textbf{Scenario} ($\mathcal{R}:\mathcal{A}$) & \textbf{FAD}  & \textbf{CLAP$_\mathcal{A}$}                 & \textbf{WER} (\%)  & \textbf{SSM}  \\ \midrule
Clean $\rightarrow$ Clean                     & 0.345         & 0.633                         & 4.8           & 0.814         \\
Clean $\rightarrow$ Env                       & 0.278         & 0.627                         & 5.8           & 0.785         \\
Env $\rightarrow$ Clean                       & 0.554         & 0.479                         & 22.6          & 0.718         \\ 
Env $\rightarrow$ Env                         & 0.276         & 0.609                         & 24.0          & 0.706         \\ 
\bottomrule
\end{tabular}
\vspace{-7pt}
\label{tab:transfer_audio}
\end{table}

\begin{table}[t]
\centering
\footnotesize
\xcaption{Objective evaluation of generated speech with simulated dataset conditioned on text reference prompts ($\mathcal{R}$: $\mathcal{T}$).}
\begin{tabular}{l|cccc}
\toprule
\textbf{Scenario} ($\mathcal{R}:\mathcal{T}$) & \textbf{FAD}  & \textbf{CLAP$_\mathcal{T}$}& \textbf{WER} (\%)  & \textbf{SSM}  \\ \midrule
Clean $\rightarrow$ Clean                     & 1.096         & 0.445                         & 6.8           & 0.779         \\
Clean $\rightarrow$ Env                       & 0.681         & 0.463                         & 7.7           & 0.701         \\
Env $\rightarrow$ Clean                       & 1.114         & 0.437                         & 31.3          & 0.654         \\ 
Env $\rightarrow$ Env                         & 0.690         & 0.446                         & 25.7          & 0.673         \\                  
\bottomrule
\end{tabular}
\label{tab:transfer_text}
\vspace{-12pt}
\end{table}

\begin{table}[t]
\centering
\footnotesize
\xcaption{Comparison of baselines and the proposed model. $\mathcal{R}$ denotes the reference prompt.}
\resizebox{\columnwidth}{!}{
\begin{tabular}{lc|cccc}
\toprule
\textbf{Model}                   & $\mathcal{R}$ & \textbf{FAD}  & \textbf{CLAP$_\mathcal{A}$} & \textbf{CLAP$_\mathcal{T}$}    & \textbf{WER} (\%)          \\ \midrule
AudioLDM                & $\mathcal{A}$ &  0.298& 0.600              &  0.349                & -            \\
                        & $\mathcal{T}$ &  0.697& 0.359              &  0.349                & -            \\ \cmidrule{2-6}
                        & Mean          &  0.498 & 0.480             &  0.349                & -            \\ \midrule
VoiceLDM                & $\mathcal{A}$ &  \textbf{0.288}& \textbf{0.643}     &  0.376                & 14.3         \\
                        & $\mathcal{T}$ &  0.753 & 0.417              &  0.361                & 12.2         \\ \cmidrule{2-6}
                        & Mean          &  0.521 & 0.530             &  0.369                & 13.3            \\ \midrule
\textbf{\shortname}     & $\mathcal{A}$ &  0.323& 0.591              &  \textbf{0.408}       & \textbf{8.6}          \\
                        & $\mathcal{T}$ &  \textbf{0.602} & \textbf{0.479}     &  \textbf{0.447}       & \textbf{11.4}         \\   \cmidrule{2-6}
                        & Mean          &  \textbf{0.463} & \textbf{0.535}             &  \textbf{0.428}                & \textbf{10.0}            \\
\bottomrule
\end{tabular}
}
\label{tab:comparison}
\vspace{-2pt}
\end{table}


\xection{Experiment Results}
In this section, we present the results of our experiments across different datasets and metrics.
Demo audio samples from our experiments are available online\footnotemark[1].

\xubsection{Performance on Simulated Recordings}
In Table~\ref{tab:transfer_audio} and~\ref{tab:transfer_text}, we report the acoustic scene transfer performances of \shortname~across diverse pairs of content and reference prompts.
\Table{transfer_audio} shows AST results using audio signals as reference prompts ($\mathcal{R}:\mathcal{A}$), while \Table{transfer_text} utilizes text descriptions generated using GPT4 for the prompt ($\mathcal{R}:\mathcal{T}$). 
Each table showcases 4 scenarios for transferring acoustic scenes: Clean $\rightarrow$ Clean, Clean $\rightarrow$ Env, Env $\rightarrow$ Clean, and Env $\rightarrow$ Env.
Here, `Env' involves environmental scenes overlaid on speech.

Our findings in \Table{transfer_audio} demonstrate that \shortname~effectively transfers acoustic scenes while maintaining spoken contents and their speakers.
However, \shortname~appears to be particularly susceptible to the `Env $\rightarrow$ Clean' transition. 
This vulnerability arises from the challenge of describing a `clean' scene, where CLAP model may not be sufficiently trained on `silence' or `quiet' acoustic environments.
In \Table{transfer_text}, we found similar results to those obtained using audio reference prompts.
Since the generated transcriptions contain complex descriptions and diverse unseen sounding objects, the results show low overall scores compared to \Table{transfer_audio}. 
While it still preserves the spoken term of content prompts well, it encounters difficulties in generating the same environment described in text prompts.
We anticipate that these issues in each table can be addressed by fine-tuning CLAP on such clean environmental datasets and utilizing more diverse text descriptions.
Additionally, providing suitable instructions or prompts to generate desirable signals through text descriptions is important.


\begin{table}[t]
\centering
\footnotesize
\vspace{-6pt}
\xcaption{Evaluation of generated speech with DDS dataset with environment reference prompts. $\mathcal{C}$ denotes the content prompt. Ground-truth MOS score is 4.206.}
\begin{tabular}{lc|cccc}
\toprule
\textbf{Model}       & $\mathcal{C}$                                      & \textbf{FAD}               & \textbf{CLAP$_\mathcal{A}$}      & \textbf{SSM}     & \textbf{MOS}    \\ \midrule
DiffRENT    & Clean                & 0.332             & 0.852                   & 0.746   & 4.000  \\
            & Env                & 0.457             & 0.792                   & 0.779   & 4.235  \\  \midrule
\textbf{\shortname}  & Clean       & 0.358             & 0.842                   & 0.723   & 3.676  \\
            & Env                & 0.478             & 0.817                   & 0.772   & 4.294  \\
\bottomrule
\end{tabular}
\label{tab:diffrent}
\vspace{-13pt}
\end{table}

\xubsection{Performance on Real Recordings}
In \Table{comparison}, we conducted a comparison between \shortname~and CLAP-based baseline models, utilizing both audio $\mathcal{A}$ and text captions $\mathcal{T}$ from the AudioCaps dataset as reference prompts $\mathcal{R}$.
Based on the FAD and CLAP scores, the proposed model demonstrates competitive generation performance compared to the baseline models.
Specifically, the signals generated by \shortname~exhibit a strong correlation with the textual descriptions (CLAP$_\mathcal{T}$) of the desired acoustic scenes.
When we measured CLAP scores between audio and text captions of the AudioCaps dataset, they scored 0.521, and the overall CLAP score of our model is the closest to the score.
Additionally, there appears to be less performance degradation when using text reference prompts compared to the baseline models.
Moreover, our model achieves better WER scores in both types of reference prompts, indicating that \shortname~effectively preserves the spoken term and accurately conveys it in the output signal.
These findings suggest that \shortname~is capable of generating speech signals accurately within a desired acoustic environment, making it well-suited for the acoustic scene transfer task.

\Table{diffrent} presents the evaluation results on DDS datasets compared with DiffRENT model, using the audio samples obtained from its demo page. 
While both models aim to achieve the same task, DiffRENT is trained exclusively on DDS datasets, which results in a better fit to the evaluation dataset.
However, our proposed model still demonstrates competitive performance, and in some metrics, it even outperforms DiffRENT.
Besides, \shortname~has superiority to learn more various audio characteristics using CLAP encoders while DiffRENT is effective in describing simple environments such as ambient noise or reverberation relying on the environmental encoder.
Also, with the help of CLAP model, \shortname~has advantages in describing in ordinary language instruction while DiffRENT always require the target audio sample.

\xection{Conclusion}

In this study, we introduced a novel task, acoustic scene transfer (AST), aimed at transforming the acoustic environments of input speech into target acoustic scenes. 
We proposed the first AST task-oriented model, \shortname, based on the latent diffusion model.
By leveraging CLAP model for acoustic scene analysis, \shortname~successfully altered and represented environments within input speech signals, effectively removing original acoustic scenes.
Experimental results have confirmed the potential of \shortname~for the AST task, demonstrating its ability to transfer acoustic scenes as intended while preserving spoken terms and voice identity in the input speech.

We believe this research would contribute significantly to the field of speech processing, offering enriched and immersive speech generation capabilities. 
For future work, we recommend focusing on more precise model implementations and refining training strategies.
Additionally, we need to explore optimal prompt provisions for generalization and enable wider and more nuanced representations of reference prompts.

\clearpage
\bibliographystyle{IEEEtran}
\bibliography{shortstrings,refs}
\end{document}